\documentstyle[aps,prl,psfig]{revtex}
\newcommand{\avg}[1]{\langle{#1}\rangle}
\newcommand{\bra}[1]{\langle{#1}|}
\newcommand{\ket}[1]{|{#1}\rangle}
\newcommand{\Avg}[1]{\left\langle{#1}\right\rangle}
\newcommand{\be}{\begin{equation}}
\newcommand{\ee}{\end{equation}}
\newcommand{\beas}{\begin{eqnarray*}}
\newcommand{\eeas}{\end{eqnarray*}}
\newcommand{\bea}{\begin{eqnarray}}
\newcommand{\eea}{\end{eqnarray}}
\newcommand{\req}[1]{(\ref{#1})}
\newcommand{\ovl}[1]{\overline{#1}}
\def\sign{\hbox{sign}\,}

\begin{document}

\twocolumn[\hsize\textwidth\columnwidth\hsize\csname
@twocolumnfalse\endcsname
\title{Continuum time limit and stationary states of the Minority Game}
\author{Matteo Marsili}
\address{Istituto Nazionale per la Fisica della Materia (INFM),
Trieste-SISSA Unit,\\
Via Beirut 2-4, Trieste 34014, Italy}
\author{Damien Challet}
\address{Theoretical Physics, Oxford University, 1 Keble Road, Oxford OX1 3NP, United Kingdom}
\date{\today}
\maketitle

\begin{abstract}
We discuss in detail the derivation of stochastic differential
equations for the continuum time limit of the Minority Game. We show
that all properties of the Minority Game can be understood by a
careful theoretical analysis of such equations. In particular, {\em
i)} we confirm that the stationary state properties are given by the
ground state configurations of a disordered (soft) spin system; {\em
ii)} we derive the full stationary state distribution; {\em iii)} we
characterize the dependence on initial conditions in the symmetric
phase and {\em iv)} we clarify the behavior of the system as a
function of the learning rate. This leaves us with a complete and
coherent picture of the collective behavior of the Minority Game.
Strikingly we find that the temperature like parameter which is
introduced in the choice behavior of {\em individual} agents turns out
to play the role, at the {\em collective} level, of the inverse of a
thermodynamic temperature.
\end{abstract}
]

\section{Introduction}

Even under the most demanding definition, the Minority Game (MG)
\cite{CZ97,web} definitely qualifies as a complex system.  The MG can
be regarded as an Ising model for systems of heterogeneous adaptive
agents, who interact {\em via} a global mechanism that entails competition 
for limited resource, as found for instance in biology and financial markets. 
 In spite of more than
three years of intense research, its rich dynamical behavior is still
the subject of investigations, many variations of the basic MG
being proposed, each uncovering new surprising regions of phase space.

Most importantly, Refs. \cite{CMZ,MCZ,MMM,MC,London} have shown that
much theoretical insight can be gained on the behavior of this class
of models, using non-equilibrium statistical physics and statistical
mechanics of disordered systems. The approach of Refs.
\cite{CMZ,MCZ,MMM,MC} rests on the assumption that, in a continuum
time limit (CTL), the dynamics of the MG can be described by a set of
{\em deterministic} equations. From these, one derives a function $H$
which is minimized along all trajectories; hence, the stationary state
of the system corresponds to the ground state of $H$, which can be
computed exactly by statistical mechanics techniques. This approach
has been challenged in Refs. \cite{Oxf1,Oxf2}, which have proposed a
{\em stochastic} dynamics for the MG, thus leading to some debate in
the literature \cite{comment,reply}.

In this paper, our aim is to analyze in detail the derivation of
the CTL in order to clarify this issue. 
We show that a proper derivation of the CTL
indeed reconciles the two approaches: the resulting dynamical
equations -- Eqs. (\ref{dupsdt31}-\ref{dupsdt33}) below, which are our
central result -- 
are indeed {\em stochastic}, as suggested in
Ref. \cite{Oxf1,Oxf2}, but still the stationary state of the dynamics
is described by the minima of the function $H$, as suggested in Refs.
\cite{CMZ,MCZ}. We then confirm the analytic results derived
previously. In few words, our analysis follows two main steps: first 
we characterize the average behavior of agents by computing the
frequency with which they play their strategies. This step can be
translated in the study of the ground state properties of a soft spin
disordered Hamiltonian. Secondly we characterize the fluctuations
around the average behavior. To do this, we explicitly solve the
Fokker-Planck equation associated to the stochastic dynamics. 

The new results which we derive are:
\begin{enumerate}
\item we derive the full probability distribution in the stationary
state. Remarkably we find that the parameter which is introduced as a
temperature in the individual choice model, turns out to play the role
of the inverse of a global temperature;
\item for $\alpha>\alpha_c$ the distribution factorizes over the
agents whereas in the symmetric phase ($\alpha<\alpha_c$) agents play
in a correlated way. In the latter case, the correlations contribute
to the stochastic force acting on agents. We show how the dependence
of global efficiency on individual temperature found in
Ref. \cite{Oxf1} arises as a consequence of these correlations;
\item we extend the analytic approach of Refs. \cite{CMZ,MCZ} to the
$\alpha<\alpha_c$ phase and asymmetric initial conditions.  The
dependence on the initial conditions in this phase, first noticed and
discussed in Refs. \cite{CMZ,MCZ}, has been more recently studied
quantitatively in Refs. \cite{Oxf2,London}. We clarify the origin of
this behavior and derive analytic solutions in the limit $\Gamma\to 0$.
\item we show that the stronger is the initial asymmetry
in agents evaluation of their strategies, the larger is the efficiency
and the more stable is the system against crowd effects\cite{Johnson}.
\item we derive the Hamiltonian of MGs with non-linear payoffs.
\end{enumerate}

This leaves us with a coherent picture of the collective behavior of
the Minority Game which is an important reference framework for the
study of complex systems of heterogeneous adaptive agents.

\section{The Model}

The dynamics of the MG is defined in terms of dynamical variables
$U_{s,i}(t)$ in discrete time $t=0,1,\ldots$ . These are scores,
propensities or ``attractions'' \cite{CamererHo} which each agent
$i=1,\ldots,N$ attaches to each of his possible choices
$s=1,\ldots,S$. Each agent takes a decision $s_i(t)$ with 

\be 
{\rm Prob}\{s_i(t)=s\}
= \frac{e^{\Gamma_i
U_{s,i}(t)}}{\sum_{s'}e^{\Gamma_i U_{s',i}(t)}}
\label{piu}
\ee

\noindent
where $\Gamma_i>0$ appears as an ``individual inverse temperature''.
The original MG corresponds to $\Gamma_i=\infty$ \cite{CZ97}, and was
generalized later to $\Gamma_i\equiv \Gamma<\infty$ \cite{Oxf1}.

The public information variable $\mu(t)$ is given to all agents; it 
belongs to the set of integers
$(1,\ldots,P)$, and can either be the binary encoding of last $M$
 winning choices \cite{CZ97}, or 
drawn at random from a uniform distribution \cite{cavagna};
 we stick to the latter case
 for sake of simplicity\footnote{Both 
prescriptions lead to qualitatively similar results for the 
quantities we study here. See \cite{CM00}
 for more details.}. The action
$a_{s_i(t),i}^{\mu(t)}$ of each agent depends the on its choice
$s_i(t)$ and on $\mu(t)$. The coefficients $a_{s,i}^\mu$, called
strategies, plays the role of quenched disorder: they are randomly
drawn signs (${\rm Prob}\{a_{s,i}^\mu=\pm 1\}=1/2$), independently for
each $i$, $s$ and $\mu$. On the basis of the outcome 

\be
A(t)=\sum_{i=1}^N a_{s_i(t),i}^{\mu(t)} 
\ee 

\noindent
each agent updates his scores according to

\be 
U_{s,i}(t+1)=U_{s,i}(t)-a_{s,i}^{\mu(t)}\frac{A(t)}{P}.
\label{updUsi}
\ee 

\noindent
The idea of this equation is that agents reward
[$U_{s,i}(t+1)>U_{s,i}(t)$] those strategy which would have predicted
the {\em minority sign} $-A(t)/|A(t)|$. The MG was initially proposed
with a nonlinear dependence on $A(t)$, i.e. with a dynamics
$U_{s,i}(t+1)=U_{s,i}(t)-a_{s,i}^{\mu(t)}{\rm sign}\,A(t)$. This leads
to qualitatively similar results. The extension of our theory to
non-linear cases is dealt with in the appendix. We shall not discuss
any longer the interpretation of the model, which is discussed at
length elsewhere \cite{MCZ,ZENews,Savit,CCMZ}.

The source of randomness are in the choices of $\mu(t)$ by Nature 
and of $s_i(t)$ by agents. These are fast fluctuating degrees of
freedom. As a consequence also $U_{s,i}(t)$ and hence the probability
with which agents chose $s_i(t)$ are subject to stochastic
fluctuations. Our analysis will indeed focus on the characterization
of the low-frequency fluctuations of $U_{s,i}$, by integrating our 
the high frequency fluctuations of $\mu(t)$ and $s_i(t)$. This will
become clearer in the next section. For the time being let it suffice
to say that there are two level of fluctuations, that of ``fast''
variables $\mu(t)$ and $s_i(t)$ and that of ``slow'' degrees of
freedom $U_{s,i}(t)$.

The key parameter is the ratio $\alpha=P/N$ \cite{Savit} and two
relevant quantities are

\be
\sigma^2=\avg{A^2},~~~~~H=\frac{1}{P}\sum_{\mu=1}^P\avg{A|\mu}^2
\label{Hdef}
\ee

\noindent
which measure, respectively, global efficiency and 
predictability\footnote{Averages $\avg{\ldots}$ stand for time averages 
in the stationary state of the process. Then $\avg{\ldots|\mu}$ stands for 
time averages conditional on $\mu(t)=\mu$.}.

Generalizations of the model, where agents account for their market
impact \cite{CMZ,MCZ}, where deterministic agents -- so-called
producers -- are present \cite{MMM}, or where agents are allowed not
to play \cite{SZ,J99,J00,BouchMG,CMZ01}, have been proposed. Rather than
dealing with the most generic model which would depend on too many
parameters, we shall limit our discussion to the plain MG. Furthermore
we shall specialize, in the second part of the paper to the case $S=2$
which lends itself to a simpler analytic treatment. The analysis
carries through in obvious ways to the more general cases discussed in
Refs. \cite{MCZ,MMM,CMZ,CMZ01}.

\section{The continuum time limit}

Our approach, which follows that of Refs. \cite{CMZ,MCZ}, is based on two key
observations:
\begin{enumerate}
\item the scaling $\sigma^2\sim N$, at fixed $\alpha$, suggests that
typically $A(t)\sim\sqrt{N}$. Hence time increments of $U_{s,i}(t)$, in 
Eq. \req{updUsi} are small (i.e. of order $\sqrt{N}/P\sim 1/\sqrt{N}$);
\item characteristic times
of the dynamics are proportional to $P$. Naively this is because
agents need to ``test'' their strategies against all $P$ values of
$\mu$, which requires of order $P$ time steps. More precisely, one can
reach this conclusion by measuring relaxation or correlation times
and verifying that they indeed grow linearly with $P$ (see Ref. 
\cite{comment}).
\end{enumerate}

The second observation implies that one needs to study the dynamics in
the rescaled time $\tau=t/P$. This makes our approach differ from that
of Refs. \cite{Oxf1,Oxf2}, where the time is not rescaled. 

In order to study the dynamics for $P,N\gg 1$ at fixed $\alpha$, we shall focus on a fixed
small increment $d\tau$ such that $Pd\tau=\alpha Nd\tau\gg 1$. This
means that we take the continuum time limit $d\tau\to 0$ {\em only
after} the thermodynamic limit $N\to\infty$. We focus only
on the leading order in $N$. Furthermore we shall also consider
$\Gamma_i$ finite and

\be
\Gamma_i d\tau\ll 1
\label{cond-Gamma}
\ee 
which means that the limit $\Gamma_i\to\infty$ should be taken
after the limit $d\tau\to 0$. The orders in which these limits are
taken, given the agreement with numerical simulation results, does not
really matters: as we shall see differences only enter in the finite
size corrections. We shall come back later to these issues.

Iteration of the dynamics for $Pd\tau$ time steps, from $t=P\tau$ to
$t=P(\tau+d\tau)$ gives

\be
u_{s,i}(\tau+d\tau)-u_{s,i}(\tau)=-
\frac{1}{P}
\sum_{t=P\tau}^{P(\tau+d\tau)-1} a_{s,i}^{\mu(t)}A(t).
\label{dupsdt}
\ee
where we have introduced the functions $u_{s,i}(\tau)=U_{s,i}(P\tau)$.

Let us separate a deterministic ($du_{s,i}$) from a stochastic 
($dW_{s,i}$) term in this equation
by replacing 

\be
a_{s,i}^{\mu(t)}A(t)=\ovl{a_{s,i}\avg{A}_\pi}+X_{s,i}(t).
\label{Xsit}
\ee
Here and henceforth, we denote averages over $\mu$ by an over-line
\[
\ovl{R}=\frac{1}{P}\sum_{\mu=1}^P R^\mu,
\]

\noindent
while $\avg{\ldots}_\pi$ stands for an average over the
distributions

\be
\pi_{s,i}(\tau)=\frac{1}{Pd\tau}\sum_{t=P\tau}^{P(\tau+d\tau)-1}
\frac{e^{\Gamma_i U_{s,i}(t)}}{\sum_{r}e^{\Gamma_i U_{r,i}(t)}},
\label{pisi}
\ee

\noindent
which is the frequency with which agent $i$ plays strategy $s$ in the
time interval $P\tau\le t<P(\tau+d\tau)$. 

Notice that $\pi_{s,i}(\tau)$ will themselves be stochastic 
variables, hence we also define the average on the stationary state as

\be
\avg{\ldots}=\lim_{\tau_0,T\to\infty}\frac{1}{T}\int_{\tau_0}^{\tau_0+T}
d\tau \avg{\ldots}_{\pi(\tau)}
\label{avgSS}
\ee
\noindent
where the average inside the integral is performed with the
probabilities $\pi_{s,i}(\tau)$.

Hence:
\bea
u_{s,i}(\tau+d\tau)&-&u_{s,i}(\tau)=du_{s,i}(\tau)+
dW_{s,i}(\tau)
\nonumber\\
&=&-\ovl{a_{s,i}\avg{A}_\pi}d\tau+
\frac{1}{P}
\sum_{t=P\tau}^{P(\tau+d\tau)-1} 
X_{s,i}(t)
\label{dupsdt2}
\eea

\noindent
Now the first term is of order $d\tau$ as required for a deterministic
term. In addition it remains finite as $N\to\infty$ \cite{NotaN}.

The second term is a sum of $Pd\tau$ random variables $X_{s,i}(t)$
with zero average. We take $d\tau$ fixed and $N$ very large, so that
$Pd\tau\gg 1$ and we can use limit theorems.  The variables
$X_{s,i}(t)$ are independent from time to time, because both $\mu(t)$
and $s_j(t)$ are drawn independently at each time. Hence $X_{s,i}(t)$
for $P\tau\le t<P(\tau+d\tau)$ are independent and
identically\footnote{Strictly speaking, $s_i(t)$ is drawn from the
distribution in Eq. \req{piu} and not from $\pi_{s,i}$ of
Eq. \req{pisi}. However these two distributions differ by a negligible
amount as long as the condition \req{cond-Gamma} holds (see later).}
distributed.  For $Pd\tau\gg 1$ we may approximate the second term
$dW_{s,i}$ of Eq. \req{dupsdt2} by a Gaussian variable with zero
average and variance 

\beas 
\Avg{dW_{s,i}(\tau)dW_{r,j}(\tau')}\!&=&\!\frac{\delta(\tau-\tau')}{P^2}
\sum_{t=P\tau}^{P(\tau+d\tau)}\!\Avg{X_{s,i}(t)X_{r,j}(t)}_\pi \\ 
&=&\delta(\tau-\tau')d\tau\frac{\Avg{X_{s,i}(t)X_{r,j}(t)}_\pi}{P} 
\eeas

\noindent
where the $\delta(\tau-\tau')$ comes from independence of $X_{s,i}(t)$
and $X_{r,j}(t')$ for $t\not = t'$ and the fact that $X_{s,i}(t)$ are
identically distributed in time leads to the expression in the second
line.  Now: 

\be 
\frac{\Avg{X_{s,i}(t)X_{r,j}(t)}_\pi}{P}=
\frac{\ovl{a_{s,i}a_{r,j}\avg{A^2}_\pi}}{P}-
\frac{\ovl{a_{s,i}\avg{A}_\pi}\,\ovl{a_{r,j}\avg{A}_\pi}}{P}
\label{XrjXsi}
\ee 

\noindent
The second term always vanishes for $N\to \infty$ because
$\ovl{a_{s,i}\avg{A}_\pi}$ is of order $N^0$ \cite{NotaN}. 
In the first term, instead,

\be
\avg{A^2|\mu}_\pi=N+\sum_{k\neq l=1}^N\sum_{s',r'}
a_{s',k}^\mu a_{r',l}^\mu\pi_{s',k}\pi_{r',l}
\label{A2}
\ee

\noindent
is of order $N$ which then gives a positive contribution in
Eq. \req{XrjXsi} for $N\to\infty$. 

Eq. \req{A2} leads to a stochastic dynamics where the noise covariance
depends on the stochastic variables $u_{s,i}(\tau)$ themselves.  The
complications which result from this fact can be avoided if one takes
the approximation

\be
\avg{A^2|\mu}_\pi\approx\ovl{\avg{A^2}}\equiv \sigma^2.
\label{approx}
\ee

\noindent
This approximation can be justified naively by observing that the
dependence on $\pi_{s,i}$ of the correlations only involves the global
quantity $\avg{A^2|\mu}_\pi/P$ for which one expects some sort of
self-averaging properties. Numerical results suggest that terms which
are ignored by Eq. \req{approx} are negligible for $N\gg 1$, but we
were unable to prove this in general \cite{thanxH} (Eq. \req{approx}
holds trivially in the limit $\Gamma\to 0$). 

\noindent
Within this approximation, the correlation, for $N\gg 1$, become
\be
\Avg{dW_{s,i}(\tau)dW_{r,j}(\tau')}\cong \frac{\sigma^2}{\alpha
N}\ovl{a_{s,i}a_{r,j}}\delta(\tau-\tau')d\tau 
\ee 

\noindent
Note that, for $r\neq s$ or $j\neq i$, correlations
$\Avg{dW_{s,i}(\tau)dW_{r,j}(\tau')}\propto\ovl{a_{s,i}a_{r,j}}\sim
1/\sqrt{N}$ vanishes as $N\to\infty$.  However it is important to
keep the off-diagonal terms because they keeps the dynamics of the phase
space point $\ket{U (t)}= \{U_{s,i}(t)\}_{s=1,\ldots,S,~i=1,\ldots,N}$
constrained to the linear space spanned by the vectors $\ket{a^\mu}=
\{a^\mu_{s,i}\}_{s=1,\ldots,S,~i=1,\ldots,N}$ which contains the
initial condition $\ket{U(0)}$. The original dynamics of $U_{s,i}(t)$
indeed posses this property.

It is important to remark that the approximation Eq. \req{approx}
makes our approach a self-consistent theory for $\sigma^2$.  We
introduce $\sigma^2$ as a constant in Eq. \req{approx} which then has
to be computed self-consistently from the dynamic equations.

\begin{figure}
\centerline{\psfig{file=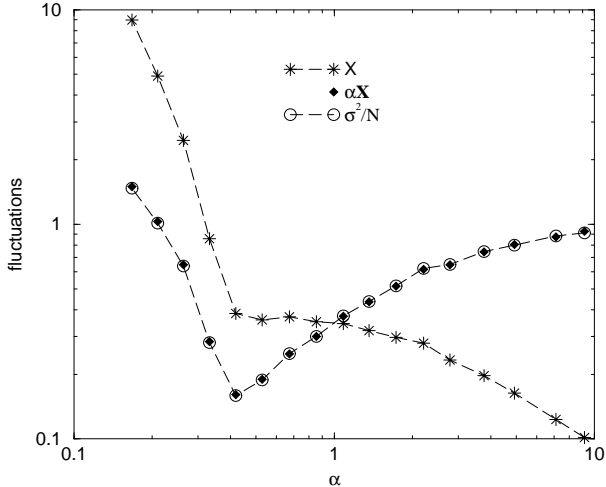,width=8cm,angle=0}}
\caption{Noise strength averaged over all agents (stars); when it is
multiplied by $\alpha$ (diamonds), one recovers $\sigma^2/N$ (circles)
($P=32$, $S=2$, $300P$, iterations, $\Gamma=\infty$, average over 50
samples). Dashed lines are for eye guidance only}
\label{figs2noise}
\end{figure}

Summarizing, the dynamics of $u_{s,i}$ is described by a
continuum time Langevin equation:
\bea
\frac{d{u}_{s,i}(\tau)}{d\tau}&=&-\ovl{a_{s,i}\avg{A}}
+\eta_{s,i}(\tau)
\label{dupsdt31}
\\
\Avg{\eta_{s,i}(\tau)}&=&0
\label{dupsdt32}
\\
\Avg{\eta_{s,i}(\tau)\eta_{r,j}(\tau')}&\cong&\frac{\sigma^2}{\alpha N}
\ovl{a_{s,i}a_{r,j}}\delta(\tau-\tau').
\label{dupsdt33}
\eea 

Equation \req{dupsdt31}, given its derivation, has to be interpreted
in the Ito sense.  The expression for the noise strength is confirmed
by figure \ref{figs2noise} where the measure of $X=\sum_{i,s}\avg{(X_{i,s}^2}/(PNS)$
in a MG is reported; note that these numerical simulations
were done for $\Gamma=\infty$ and confirm Eq. \req{dupsdt33} is valid
even for $\alpha<\alpha_c$. Fig. \ref{figs2gamma} compares the results
of numerical simulations of the MG, as a function of $\Gamma$, with
those of a semi-analytic solution of Eqs.
(\ref{dupsdt31}-\ref{dupsdt33}), to be discussed later. The agreement
of the two approaches shows that Eqs. (\ref{dupsdt31}-\ref{dupsdt33})
are valid even in the symmetric phase ($\alpha\le\alpha_c$) for all
values of $\Gamma$. 

The instantaneous probability distribution of $s$, in the continuum
time limit, reads 
\be
\pi_{s,i}(\tau)=\frac{e^{\Gamma_i u_{s,i}(\tau)}}{\sum_r
e^{\Gamma_i u_{r,i}(\tau)}}. 
\label{pisi1}
\ee

This and Ito calculus then lead to a
dynamic equation for $\pi_{s,i}(\tau)$. We prefer to exhibit this for
$\Gamma_i=\Gamma$ and using the rescaled time $t=\Gamma \tau$:

\bea
\frac{d\pi_{s,i}}{dt}&=&-\pi_{s,i}\left[\ovl{a_{s,i}\avg{A}}-
\vec{\pi}_i\cdot\ovl{\vec{a}_i\avg{A}}\right] \nonumber\\ 
&+&
\frac{\sigma^2\Gamma}{\alpha
N}\pi_{s,i}\left(\pi_{s,i}-\vec{\pi}_i^2\right)
+\sqrt{\Gamma}\pi_{s,i}\left(\eta_{s,i}-\vec{\pi}_i\vec{\eta}_i\right).
\label{dynpi}
\eea 
The first term in the r.h.s. comes from the deterministic part of
Eq. \req{dupsdt31}, the second from the Ito term (where we neglected
terms proportional to $\ovl{a_{i,s}a_{i,s'}}\sim 1/\sqrt{N}$ for $s\ne
s'$) and the third from the stochastic part. It is clear that, in the
limit $\Gamma\to 0$ the last two term vanish and the dynamics
becomes deterministic.

We see then that $\Gamma$ tunes the strength of stochastic
fluctuations in much the same way as temperature does for thermal
fluctuations in statistical mechanics. The ``individual inverse
temperature'' $\Gamma$ should indeed more correctly be interpreted as
a learning rate\footnote{This is an {\em a posteriori} learning
rate. Indeed $1/\Gamma$ is the time the dynamics of the scores needs
in order to learn a payoff difference. From a different viewpoint,
$\Gamma$ tunes the randomness of the response of agents. The larger the
randomness, the longer it takes to average fluctuations out.}.
Furthermore it plays the role of a ``global temperature''. We shall
pursue this discussion in detail below for the case $S=2$.


At this point, let us comment on the limit $\Gamma\to\infty$, which
is of particular importance, since it corresponds to the original MG.
It is clear that in
the limit $\Gamma\to\infty$ the dynamical equations \req{dynpi} become
problematic. The origin of the problem lies in the order in which the
limits $N\to\infty$ and $\Gamma\to\infty$ is performed. Indeed in
Eq. \req{pisi} $U_{s,i}(t)\simeq u_{s,i}(\tau)+O(d\tau)$ for $P\tau\le
t< P(\tau+d\tau)$. Therefore, as long as $\Gamma d\tau\ll 1$ the
difference between Eq. \req{pisi} and $\pi_{s,i}$ in Eq. \req{pisi1}
is negligible. In practice, in order to satisfy both $\Gamma d\tau\ll
1$ and $P d\tau\gg 1$, one needs $\Gamma\ll P$. When this condition
is not satisfied the instantaneous probability Eq. \req{piu}
fluctuates very rapidly at each time-step. Eq. \req{pisi} averages out
these high frequency fluctuations so that, even for $\Gamma=\infty$,
the distribution $\pi_{s,i}(\tau)$ of Eq. \req{pisi} is not a
discontinuous step function of $u_{s,i}(\tau)$, as suggested by
Eq. \req{pisi1}. High frequency fluctuations contribute to the
functional form of $\pi_{s,i}$ on $u_{s,i}$ which will differ from
Eq. \req{pisi1}. 

Summarizing, when we let $\Gamma\to\infty$ only after the limit
$N\to\infty$ has been taken no problem arises. There is no reason to
believe that results change if the order of the limits is
interchanged. This expectations, as we shall see, is confirmed by
numerical simulations (see Fig. \ref{figs2gamma}): direct numerical
simulations of the MG deviate from the prediction of
Eqs. (\ref{dupsdt31}-\ref{dupsdt33}) only for finite size effects
which vanish as $N\to\infty$.


Eqs. (\ref{dupsdt31}-\ref{dupsdt33}) are our central result.  We shall
devote the rest of the paper to discuss their content and to show that
all of the observed behavior of the MG can be derived from these
equations. 

\section{Stationary state}

Let us take the average, denoted by $\avg{\ldots}$, of
Eq. \req{dupsdt31} on the stationary state (SS).  Let
\[
f_{s,i}=\avg{\pi_{s,i}}
\]
be the frequency with which agent $i$ plays strategy $s$ in the SS.
Then we have
\[
v_{s,i}\equiv\Avg{\frac{d U_{s,i}}{d\tau}}=-\ovl{a_{s,i}\avg{A}},~~~~~~
\avg{A|\mu}=\sum_{j,s'}f_{s',j}a_{s',j}^\mu
\]
Given the relation between $\pi_{s,i}$ and $U_{s,i}$ and considering that
the long time dynamics of $U_{s,i}$ in the SS is $U_{s,i}(\tau)={\rm const}+
v_{s,i}\tau$, we have that 
{\em i)} each strategy which is played in the SS by agent 
$i$ must have the same ``velocity'' $v_{s,i}=v_i^*$, and {\em ii)} 
strategies which
are not played (i.e. with $f_{s,i}=0$) must have $v_{s,i}<v_i^*$. In other
words

\bea
-\ovl{a_{s,i}\avg{A}}&=&v_i^*,~~~\forall i,s ~~\hbox{such that}~~f_{s,i}>0
\label{SS1}\\
-\ovl{a_{s,i}\avg{A}}&\le&v_i^*,~~~\forall i,s ~~\hbox{such that}~~f_{s,i}=0.
\label{SS2}
\eea

Consider now the problem of constrained minimization of $H$ in
Eq. \req{Hdef}, subject to $f_{s,i}\ge 0$ for all $s,i$ and the
normalization conditions.  Introducing Lagrange multipliers
$\lambda_i$ to enforce $\sum_s f_{s,i}=1$ for all $i$, this problem
reads

\be
\min_{\{f_{s,i}\ge 0\}} \left\{\ovl{\avg{A}^2}-\sum_{i=1}^N \lambda_i
\left(1-\sum_{s=1}^S f_{s,i}\right)\right\}.
\label{MinH}
\ee

Taking derivatives, we find that if $f_{s,i}>0$ then
$\ovl{a_{s,i}\avg{A}}+\lambda_i=0$ whereas if $f_{s,i}=0$ then
$\ovl{a_{s,i}\avg{A}}+\lambda_i\ge 0$. These are exactly
Eqs. (\ref{SS1},\ref{SS2}) where $v_i^*=\lambda_i$. We then conclude that
the two problems Eqs. (\ref{SS1},\ref{SS2}) and Eq. \req{MinH} are one and
the same problem\footnote{Indeed both problems can be put in the form of a 
Linear Complementarity problem \cite{LCP}:
\beas
f_{s,i}&\ge & 0
\\
\sum_{j,s'}\ovl{a_{s,i}a_{s',j}}f_{s',j}+v_i^*&\ge & 0
\\
f_{s,i}\left[\sum_{j,s'}\ovl{a_{s,i}a_{s',j}}f_{s',j}+v_i^*\right]
&=&0
\eeas 
This problem has a solution for all values of $v_i^*$ because of
non-negativity of the matrix $\ovl{a_{s,i}a_{s',j}}$, see
Ref. \cite{LCP}.}. In other words $f_{s,i}$ can be computed from the
constrained minimization of $H$ as proposed in Refs. \cite{CMZ,MCZ}.

Hence the statistical mechanics approach based on the study of the
ground state of $H$ is correct. This approach gives the frequency 
$f_{s,i}$ with which agents play their strategies. 

We remark once more that $H$ is a function of the stationary state 
probabilities $f_{s,i}$. Also note that 

\[
\tilde H\{\pi_{s,i}\}=\sum_{i,j=1}^N\sum_{s,s'=1}^S
\ovl{a_{s,i}a_{s',j}}\pi_{s,i}\pi_{s',j}
\]
as a function of the instantaneous
probabilities $\pi_{s,i}$, is {\em not} a Lyapunov function of the
dynamics. The dynamical variables $\pi_{s,i}(t)$ are subject to
stochastic fluctuations of the order of $\sqrt{\Gamma_i}$ around their
average values $f_{s,i}$. Only in the limit $\Gamma_i\to 0$, when the
dynamics becomes deterministic and $\pi_{s,i}\to f_{s,i}$, the quantity
$\tilde H\{\pi_{s,i}\}$ becomes a Lyapunov function. 

The solution to the minimization of $H$ reveals two qualitatively
distinct phases \cite{CMZ,MCZ} which are separated by a phase
transition occurring as $\alpha\to\alpha_c$. We discuss qualitatively 
the behavior of the solution for a generic $S$ and leave for the next 
section a more detailed discussion in the simpler case $S=2$.

\subsection{Independence on $\Gamma$ for $\alpha>\alpha_c$}

When $\alpha>\alpha_c$ the solution to Eq. \req{MinH} is unique and
$H>0$. Hence $f_{s,i}$ does not depends on $\Gamma$, neither does
$H$. In addition we shall see that 

\be
\avg{\pi_{s,i}\pi_{s',j}}=\avg{\pi_{s,i}}\avg{\pi_{s',j}}=f_{s,i}f_{s',j}
~~~\hbox{for $i\neq j$}
\label{factor}
\ee 
implying that 
\[
\sigma^2\equiv N+
\sum_{i\neq j}\sum_{s,r}\ovl{a_{s,i}a_{r,j}}\avg{\pi_{s,i}\pi_{r,j}}
\]
does not depend on $\Gamma$ either.  Hence
the solution $\{f_{s,i}\}$ uniquely determines all quantities in the
SS, as well as the parameters which enter into the dynamics (notice
the dependence on $\sigma^2$ in Eq. \ref{dupsdt33}). In particular
$\sigma^2$ does not depend on $\Gamma$.

\subsection{Dependence on $\Gamma$ and on initial conditions for 
$\alpha<\alpha_c$}

For $\alpha<\alpha_c$ the solution to the minimization problem is not
unique: there is a connected set of points $\{f_{s,i}\}$ such that
$H=0$. Let us first discuss the behavior of the system in the limit
$\Gamma\to 0$, where the dynamics becomes deterministic.  The dynamics
reaches a stationary state $\{f_{s,i}\}$ which depends on the initial
conditions. 

In order to see this, let us introduce the vector notation
$\ket{v}=\{v_{s,i},\,s=1,\ldots,S,\,i=1,\ldots,N\}$. Then for all
times $\ket{u(\tau)}$ is of the form
\[
\ket{u(\tau)}=\ket{u(0)}+\sum_{\mu=1}^P\ket{a^\mu}C^\mu(\tau)
\]
where $C^\mu(\tau)$ are $P$ functions of time.

If there are vectors $\bra{v}$ such that $\avg{v|a^\mu}=0$ for all
$\mu$, then $\avg{v|u(\tau)}=\avg{v|u(0)}$, i.e. the components of the
scores will not change at all along these vectors. As a result the SS
will depend on initial conditions $\ket{u(0)}$. These vectors
$\bra{v}$ exist exactly for $\alpha<\alpha_c$ \cite{MCZ}, because 
the ``dimensionality'' of the vectors $\ket{u(\tau)}$ is
larger than $P$\footnote{In order to compute the dimensionality of 
the vectors $\ket{u}$ we have to take into account the $N$ normalization
conditions and the fact that strategies which are not played
($f_{s,i}=0$) should not be counted. So if there are $N_>$ variables
$f_{s,i}>0$, the relevant dimension of the space of $\ket{u}$ is
$N_>-N$. Hence vectors $\bra{v}$ orthogonal to all $\ket{a^\mu}$ exist
for $N_>-N>P$, i.e. for $\alpha<\alpha_c=N_>(\alpha_c)/N-1$.}. 

The picture is made even more complex by the fact that, for
$\alpha<\alpha_c$, when $\Gamma$ is finite Eq. \req{factor} does not
hold. Hence $\sigma^2$ has a contribution which depends on the
stochastic fluctuations around $f_{s,i}$. The strength of these 
fluctuations, by Eqs. (\ref{dupsdt33},\ref{dynpi}), depends on $\Gamma$ 
and $\sigma^2$ itself. We face, in this case,
a self-consistent problem: $\sigma^2$ enters as a parameter of the
dynamics but should be computed in the stationary state of the dynamics
itself. Therefore the solution to this problem and hence $\sigma^2$ 
depends on $\Gamma$. The solution $\{f_{s,i}\}$ to the minimization
of $H$ should also be computed self-consistently. As a result, 
the SS properties acquire a dependence on $\Gamma$.

The condition \req{factor}, which is similar to the {\em clustering}
property in spin glasses~\cite{MPV}, plays then a crucial role. We
show below how the condition \req{factor}, the dependence on initial
conditions and on $\Gamma$ enter into the detailed solution for
$S=2$. By similar arguments our conclusion can be generalized to all
$S>2$.

\section{The case $S=2$}

We work in this section with the simpler case of $S=2$ strategies,
labelled by $s=\pm$. We also set $\Gamma_i=\Gamma$ for all $i$. 
Following Refs. \cite{CMZ,CM} we introduce the variables
\[
\xi_i^\mu=\frac{a_{+,i}^\mu-a_{-,i}^\mu}{2},
~~~\Omega^\mu=\sum_{i=1}^N\frac{a_{+,i}^\mu+a_{-,i}^\mu}{2}.
\]

Let us rescale time $t=\Gamma\tau$ and introduce the variables
\[
y_i(t)=\Gamma\ \frac{U_{+,i}(\tau)-U_{-,i}(\tau)}{2}
\]
Then, using Eq. \req{pisi}, the dynamical equations
(\ref{dupsdt31}-\ref{dupsdt33}) become 
\bea
\frac{dy_i}{dt}&=&-\ovl{\xi_i\Omega}-\sum_{j=1}^N\ovl{\xi_i\xi_j}\tanh
(y_j) +\zeta_i
\label{langS21}
\\
\avg{\zeta_i(t)\zeta_j(t')}&=&\frac{\Gamma\sigma^2}{\alpha N}
\ovl{\xi_i\xi_j}\delta(t-t').
\label{langS22}
\eea

The Fokker-Planck (FP) equation for the probability distribution 
$P(\{y_i\},t)$ under this dynamics reads 
\bea
\frac{\partial P(\{y_i\},t)}{\partial t}=\sum_{i=1}^N
\frac{\partial }{\partial y_i}&&\left\{\ovl{\xi_i\Omega}+
\sum_{j=1}^N\ovl{\xi_i\xi_j}\tanh (y_j)+\right.\nonumber
\\
&~&~~\left.\frac{1}{\beta}\sum_{j=1}^N
\ovl{\xi_i\xi_j}\frac{\partial }{\partial y_j}\right\}P(\{y_i\},t)
\label{FP}
\eea
where we have introduced the parameter
\be
\beta=\frac{2 \alpha N}{\Gamma\sigma^2}.
\label{beta}
\ee

Multiply Eq. \req{FP} by $y_i$ and integrate over all variables. Using
integration by parts, assuming that $P\to 0$ fast as $y_j\to\infty$,
one gets
\[
\frac{\partial }{\partial t}\avg{y_i}=-\ovl{\xi_i\Omega}-
\sum_{j=1}^N\ovl{\xi_i\xi_j}\Avg{\tanh (y_j)}.
\]

Let us look for solutions with $\avg{y_i}\sim v_i t$ and define
$m_i=\avg{\tanh (y_i)}$. Hence for
$t\to\infty$ we have
\be
v_i=-\ovl{\xi_i\Omega}-
\sum_{j=1}^N\ovl{\xi_i\xi_j}m_j.
\label{statcond}
\ee

\noindent
Now, either $v_i=0$ and $\avg{y_i}$ is finite, or 
$v_i\neq 0$, which means that $y_i\to \pm\infty$
and $m_i={\rm sign}\, v_i$. In the latter case ($v_i\neq 0$) 
we say that agent $i$ is {\em frozen} \cite{CM},
we call ${\cal F}$ the set of frozen agents and $\phi=|{\cal F}|/N$ 
the fraction of frozen agents.

As in the general case, the parameters $v_i$ for $i\in {\cal F}$ and
$m_i\equiv\avg{\tanh (y_i)}$ for $i\not\in {\cal F}$ are obtained by
solving the constrained minimization of
\[
H=\frac{1}{P}\sum_{\mu=1}^P
\left[\Omega^\mu+\sum_{i=1}^N\xi_i^\mu m_i\right]^2.
\]

When the solution of $\min H$ is unique, i.e. for $\alpha>\alpha_c$,
the parameters $m_i$ depend 
only on the realization of disorder $\{\xi_i^\mu,\Omega^\mu\}$, and
their distribution can be computed as in Ref \cite{CMZ}. When
the solution is not unique, i.e. for $\alpha<\alpha_c$, we are
left with the problem of finding which solution the dynamics selects.
Let us suppose that we have solved this problem (we shall come back
later to this issue), so that all $m_i$ are known.

Using the stationary condition Eq. \req{statcond}, we can write 
the FP equation for the probability distribution
$P_u(y_i,i\not\in{\cal F})$ of unfrozen agents. For times
so large that all agents in ${\cal F}$ are indeed frozen 
(i.e. $s_i(t)={\rm sign}\, v_i$) this reads:
\[
\frac{\partial P_u}{\partial t}=\sum_{i\not\in {\cal F}}
\frac{\partial }{\partial y_i}
\sum_{j\not\in {\cal F}}\ovl{\xi_i\xi_j}
\left\{\tanh (y_j)-m_j+\frac{1}{\beta}
\frac{\partial }{\partial y_j}\right\}P_u.
\]
This has a solution
\be
P_u\propto \exp\left\{-\beta\sum_{j\not\in{\cal F}}
\left[\log\cosh y_j-m_j y_j\right]\right\}.
\label{sol1}
\ee

Finally we have to impose the constraint that
$\ket{y(t)}=\{y_i(t)\}_{i=1}^N$ must lie on the linear space spanned
by the vectors $\ket{\xi^\mu}$ which contains the initial condition
$\ket{y(0)}$. This means that
\be
P_u\propto 
{\cal P}_{y(0)}
\exp\left\{-\beta\sum_{j\not\in{\cal F}}
\left[\log\cosh y_j-m_j y_j\right]\right\}
\label{sol2}
\ee
where the projector ${\cal P}_{y(0)}$ is given by
\be
{\cal P}_{y(0)}\equiv
\prod_{\mu=1}^P \int_{-\infty}^\infty dc^\mu
\prod_{i=1}^N\delta\left[y_i-y_i(0)-\sum_{\mu=1}^Pc^\mu \xi_i^\mu\right].
\label{proj}
\ee

We find it remarkable that $\Gamma$, which is introduced as the
inverse of an {\em individual} ``temperature'' in the definition of
the model, actually turns out to be proportional to $\beta^{-1}$ (see
Eq. \ref{beta}) which plays {\em collectively} a role quite similar to
that of temperature.

Using the distribution Eq. \req{sol2}, we can compute 
\bea
\sigma^2&=&H+\sum_{i=1}^N\ovl{\xi_i^2}(1-m_i^2)+\nonumber\\
&~&\sum_{i\neq j}\ovl{\xi_i\xi_j}
\avg{(\tanh y_i-m_i)(\tanh y_j-m_j)}.
\label{s2f}
\eea
This depends on $\beta$, i.e. on $\sigma^2$ itself by virtue of
Eq. \req{beta}. The stationary state is then the solution of a 
self-consistent problem.
Let us analyze in detail the solution of this self-consistent problem.

\subsection{$\alpha>\alpha_c$}

For $\alpha>\alpha_c$ the solution of $\min H$ is unique, and hence
$m_i$ depends only on the realization of the disorder. In addition,
the number $N-|{\cal F}|\equiv N(1-\phi)$ of unfrozen agents is less
than $P$ and the constraint is ineffective, i.e.  ${\cal
P}_{y(0)}\equiv 1$. The scores $\ket{y}$ of unfrozen agents span a
linear space which is embedded in the one spanned by the vectors
$\ket{\xi^\mu}$. Hence the dependence on initial conditions $y_i(0)$
drops out. Therefore the probability distribution of $y_i$ factorizes,
as in Eq. \req{sol1}. Then the third term of Eq. \req{s2f}, which is
the only one which depends on $\beta$, vanishes identically. We
conclude that, for $\alpha>\alpha_c$, $\sigma^2$ only depends on $m_i$
and is hence independent of $\Gamma$ as confirmed by numerical
simulations.

Summarizing, for $\alpha>\alpha_c$ one derives a complete solution of
the MG by finding first the minimum $\{m_i\}$ of $H$ and then by
computing $\sigma^2$, $\beta$ and the full distribution of $y_i$ from
Eq. \req{sol1}.

\subsection{$\alpha\le\alpha_c$: dependence on $\Gamma$ and crowd effects}

When $\alpha<1-\phi$, on the other hand, the solution of $\min H$ is
not unique. Furthermore the constraint cannot be integrated out and the
stationary distribution depends on the initial conditions. 

\begin{figure}
\centerline{\psfig{file=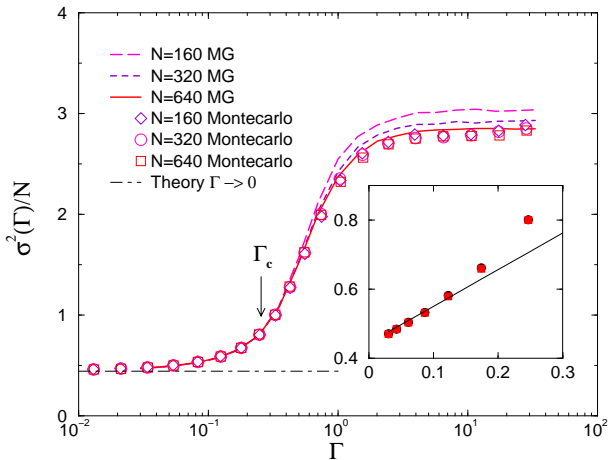,width=8cm,angle=0}}
\caption{Global efficiency $\sigma^2/N$ versus $\Gamma$ for
$\alpha=0.1<\alpha_c$ and different system sizes. Lines refer to
direct simulations of the MG with $N=160,320$ and $640$. Finite size
effect for $\Gamma\gg 1$ are evident. Symbols refer instead to the
solution of the self-consistent equation \req{self-cons} for the same
system sizes. For both methods and all values of $N$, $\sigma^2$ is
averaged over 100 realizations of the disorder.  The arrow marks the
location of $\Gamma_c$ predicted by Eq. \req{Gammac}. In the inset,
the theoretical prediction Eq. \req{s2approx} on the leading behavior
of $\sigma^2/N$ for $\Gamma\ll 1$ (solid line) is tested against
numerical simulations of the MG (points) for the same values of $N$.}
\label{figs2gamma}
\end{figure}

Numerical
simulations \cite{Oxf1} show that $\sigma^2$ increases with $\Gamma$
for $\alpha<\alpha_c$ (see Fig. \ref{figs2gamma}).  This effect has
been related to crowd effects in financial markets \cite{Johnson}.
Ref. \cite{MC} has shown that crowd effects can be fully understood in
the limit $\alpha\to 0$: as $\Gamma$ exceeds a critical learning rate
$\Gamma_c$, the time independent SS becomes unstable and a {\em
bifurcation} to a period two orbit occurs. Neglecting the stochastic
term $\zeta_i$, Ref.\cite{MC} shows that this picture can be extended
to $\alpha>0$\footnote{The idea of Ref. \cite{MC} is the following:
imagine that our system is close to a SS point $y_i^*$ at time
$t=t_k$, when $\mu(t)=1$.  Will the system be close to $y_i^*$ when
the pattern $\mu=1$ occurs again the next time $t'=t_{k+1}$?  To see
this, let us integrate Eq. \req{langS21} from $t=t_k$ to $t_{k+1}$. In
doing this we {\em i)} neglect the noise term (i.e. $\zeta_i=0$) and
{\em ii)} assume that $\tanh y_i(t)\approx\tanh y_i(t_k)$ stays
constant in the integration time interval. This latter assumption is
similar to the recently introduced \cite{London} batch version of the
MG, where agents update their strategies every $P$ time-steps. This
leads to study a discrete time dynamical system

\be
y_i(t_{k+1})=y_i(t_k)-\Gamma\left[
\ovl{\Omega\xi_i}+\sum_{j=1}^N\ovl{\xi_i\xi_j}\tanh[y_j(t_k)]
\right]
\label{dynsys}
\ee

\noindent
where the factor $\Gamma$ comes because $t_{k+1}-t_k$ is on average
equal to $\Gamma$.  The linear stability of fixed point solutions is
analyzed setting $y_i(t_k)=y^*_i+\delta y_i(k)$ and computing the
eigenvalues of the linearized map $\delta y_i(k+1)\simeq \sum_j
T_{i,j}\delta y_j(k)$.  There is a critical value of $\Gamma$ above
which the solution $y_i^*$ become unstable, which is given by
Eq. \req{Gammac}.}. This approach suggests a crossover to a ``turbulent''
dynamics for $\Gamma>\Gamma_c$, where

\be
\Gamma_c(\alpha)=\frac{4}{(1+\sqrt{\alpha})^2(1-Q)}
\label{Gammac}
\ee

\noindent
and
\[
Q=\frac{1}{N}\sum_{i=1}^N m_i^2.
\]

Both $Q$ and $\Gamma_c$ can be computed exactly in the limit
$N\to\infty$ within the statistical mechanics approach \cite{CMZ,MC}.

This approach however {\em i)} does not properly takes into account
the stochastic term, {\em ii)} does not explain what happens for
$\Gamma>\Gamma_c$ and {\em iii)} does not explains why such effects
occur only for $\alpha<\alpha_c$.

The stochastic dynamics derived previously gives detailed answers to
all these issues. We first restrict attention to symmetric initial
conditions $y_i(0)=0$ and then discuss the dependence on initial
conditions. The choice of $y_i(0)=0$ $\forall i$ is convenient because
it allows one to use this same symmetry to identify the solution
$\{m_i\}$ to the minimization of $H$, independently of $\Gamma$. 
To be more precise, one can introduce a ``magnetization'' 

\be
M\equiv
\lim_{t\to\infty}\frac{1}{N}\sum_{i=1}^N \avg{s_i(t)s_i(0)}
=\frac{1}{N}\sum_{i=1}^N m_i
\label{eqM}
\ee

which measures the overlap of the SS configuration with the initial 
condition. Symmetric initial conditions are related to $M=0$ SS. These 
are the states we focus on.
The solution is derived in two steps:
\begin{enumerate}
\item find the minimum $\{m_i\}$ of $H$, with $M=0$; 
\item compute self-consistently $\sigma^2$.
\end{enumerate}
The numerical procedure for solving the problem is
the following: given the realization of disorder $\{\xi_i^\mu$,
$\Omega^\mu\}$,  step (1) --- finding the minimum $\{m_i\}$ of $H$ --- is straightforward. For step (2) we sample the distribution
Eq. \req{sol2} with the Montecarlo method\footnote{The Montecarlo
procedure follows the usual basic steps: {\em i)} A move $y_i\to
y_i+\epsilon \xi_i^\mu$ is proposed, with $\mu$ and $\epsilon$ drawn
at random, {\em ii)} the ``energy''
\[
E\{y_i\}=\sum_{i=1}^N \left[\log\cosh y_i-m_i y_i\right]
\]
of the new configuration is computed and {\em iii)} The move is
accepted with a probability equal to $\min(1,e^{-\beta\Delta E})$
where $\Delta E$ is the ``energy'' difference.} at inverse temperature
$\beta$ and measure the $\beta$ dependent contribution of $\sigma^2$
in Eq. \req{s2f}:
\[
\Sigma(\beta)=\sum_{i\neq j}\ovl{\xi_i\xi_j}
\avg{(\tanh y_i-m_i)(\tanh y_j-m_j)}_\beta.
\]
Here $\avg{\ldots}_\beta$ stands for an average over the distribution 
Eq. \req{sol2} with parameter $\beta$. Finally we solve the equation
\be
\sigma^2(\Gamma)=\sigma^2(0)+\Sigma\left(\frac{2\alpha N}
{\Gamma\sigma^2(\Gamma)}\right).
\label{self-cons}
\ee

This procedure was carried out for different system sizes and several
values of $\Gamma$. The results, shown in Fig. \ref{figs2gamma}, agree
perfectly well with direct numerical simulations of the MG. Actually
Fig. \ref{figs2gamma} shows that, for $\Gamma\gg 1$, the solution of
the self-consistent equation \req{self-cons} suffers much less of
finite size effects than the direct numerical simulations of the MG.
Fig. \ref{figs2gamma} also shows that, even if only approximate, 
Eq. \req{Gammac} provides an useful estimate of the point where 
the crossover occurs.

It is possible to compute $\sigma^2(\Gamma)$ to leading order in
$\Gamma\ll 1$. The calculation is carried out in the appendix in
detail. The result is
\be
\frac{\sigma^2}{N}\cong\frac{1-Q}{2}\left[1+
\frac{1-Q+\alpha(1-3Q)}{4\alpha}\Gamma
+ O(\Gamma^2)\right].
\label{s2approx}
\ee 
The inset of Fig. \ref{figs2gamma} shows that this expression
indeed reproduces quite accurately the small $\Gamma$ behavior of
$\sigma^2$.  Note finally that Eq. \req{self-cons} has a finite
solution $\sigma^2(\infty)=\sigma^2(0)+\Sigma(0)$ in the limit
$\Gamma\to\infty$. Furthermore it is easy to understand the origin of
the behavior $\sigma^2/N\sim 1/\alpha$ for Eq. \req{self-cons}.
Because of the constraint, when $\ovl{\xi_i\xi_j}$ is positive
(negative) the fluctuations of $\tanh (y_i)-m_i$ are positively
(negatively) correlated with $\tanh (y_j)-m_j$. If we assume that
$\avg{[\tanh (y_i)-m_i][\tanh (y_j)-m_j]}\simeq c \ovl{\xi_i\xi_j}$ 
for some constant $c$, we
find $\Sigma\simeq c\sum_{i\neq j}\ovl{\xi_i\xi_j}^2$. This
leads easily to $\Sigma/N\simeq c/(4\alpha)$, which explains the
divergence of $\sigma^2/N$ as $\alpha\to 0$ for $\Gamma\gg 1$.

\subsection{Selection of different initial conditions in the 
Replica calculation}

As discussed above, the stationary state properties of the MG in the
symmetric phase depend on the initial conditions. Can the statistical
mechanics approach to the MG
\cite{CMZ,MCZ} be extended to characterize this dependence for
$\Gamma\ll 1$? If this is possible, how do we expect the resulting 
picture to change when $\Gamma$ increases? We shall first focus on
the first question (i.e. $\Gamma\ll 1$) and then discuss the second.

Of course one can introduce the constraint on the distribution of
$y_i$ in the replica approach in a straightforward manner. This leads
however to tedious calculations. We prefer to follow a different
approach. In the $\alpha<\alpha_c$ phase the minimum of $H$ is
degenerate, i.e. $H=0$ occurs on a connected set of points. Each of
these points corresponds to a different set of initial conditions, as
discussed above. In order to select a particular point with $H=0$
we can add a potential $\eta\sum_i(s_i-s_i^*)^2/2$ to
the Hamiltonian $H$, which will favor the solutions closer to $s_i^*$, and
then let the strength $\eta$ of the potential go to zero. This
procedure lifts the degeneracy and gives us the statistical features
of the equilibrium close to $s_i^*$. 

The nature of the stationary state changes as the asymmetry in
the initial conditions changes. If we take $s_i^*=s^*$, the 
state at $s^*=0$ describes symmetric initial conditions and
increasing $s^*>0$ gives asymmetric states.

The saddle point equations of the statistical mechanics approach of
Ref. \cite{CMZ} can be reduced to two equations:

\bea 
Q&=&\int_{-\infty}^\infty Dz s_0^2(z)
\label{sad1}\\
\chi&=&\frac{1+\chi}{\sqrt{\alpha(1+Q)}}\int_{-\infty}^\infty Dz z s_0(z)
\label{sad2}
\eea 

\noindent 
where $Dz=\frac{dz}{\sqrt{2\pi}}e^{-z^2/2}$, $s_0(z)\in
[-1,1]$ is the value of $s$ which minimizes \be
V_z(s)=\frac{1}{2}s^2-\sqrt{\frac{1+Q}{\alpha}}z
s+\frac{1}{2}\eta(1+\chi) (s-s^*)^2 \ee and $\chi=\beta(Q-q)/\alpha$
is a ``spin susceptibility''.  There are two possible solutions: one
with $\chi<\infty$ finite as $\eta\to 0$ which describes the
$\alpha>\alpha_c$ phase.  The other has $\chi\sim 1/\eta$ which
diverges as $\eta\to 0$. This solution describes the $\alpha<\alpha_c$
phase. We focus on this second solution, which can be conveniently 
parameterized by two parameters $z_0$ and $\epsilon_0$. We find 
\[
s_0(z)=\left\{
\begin{array}{cc}
-1 & \hbox{if $z\le -z_0-\epsilon_0$}\cr
\frac{z+\epsilon_0}{z_0} & \hbox{if $-z_0-\epsilon_0<z<z_0-\epsilon_0$}\cr
1 & \hbox{if $z\ge z_0-\epsilon_0$}
\end{array}
\right.
\]
Indeed Eq. \req{sad1} gives $Q(z_0,\epsilon_0)$ and Eq. \req{sad2}
which for $\chi\to\infty$ reads $\sqrt{\alpha(1+Q)}=\int Dz z s_0(z)$,
then gives $\alpha(z_0,\epsilon_0)$. 

With $\epsilon_0\neq 0$ one finds solutions with a non-zero
``magnetization'' $M=\avg{s_i}$. This quantity is particularly
meaningful, in this context, because it measures the overlap of the
behavior of agents in the SS with their a priori preferred strategies

\be
M\equiv\int_{-\infty}^\infty \!Dz\,s_0(z)
=\lim_{t\to\infty}\frac{1}{N}\sum_{i=1}^N \avg{s_i(t)s_i(0)}.
\label{eqM1}
\ee

\noindent
Note indeed that one can always perform a ``gauge'' transformation in
order to redefine $s=+1$ as the initially preferred strategy. This
amounts to taking $y_i(0)\ge 0$ for all $i$.

\begin{figure}
\centerline{\psfig{file=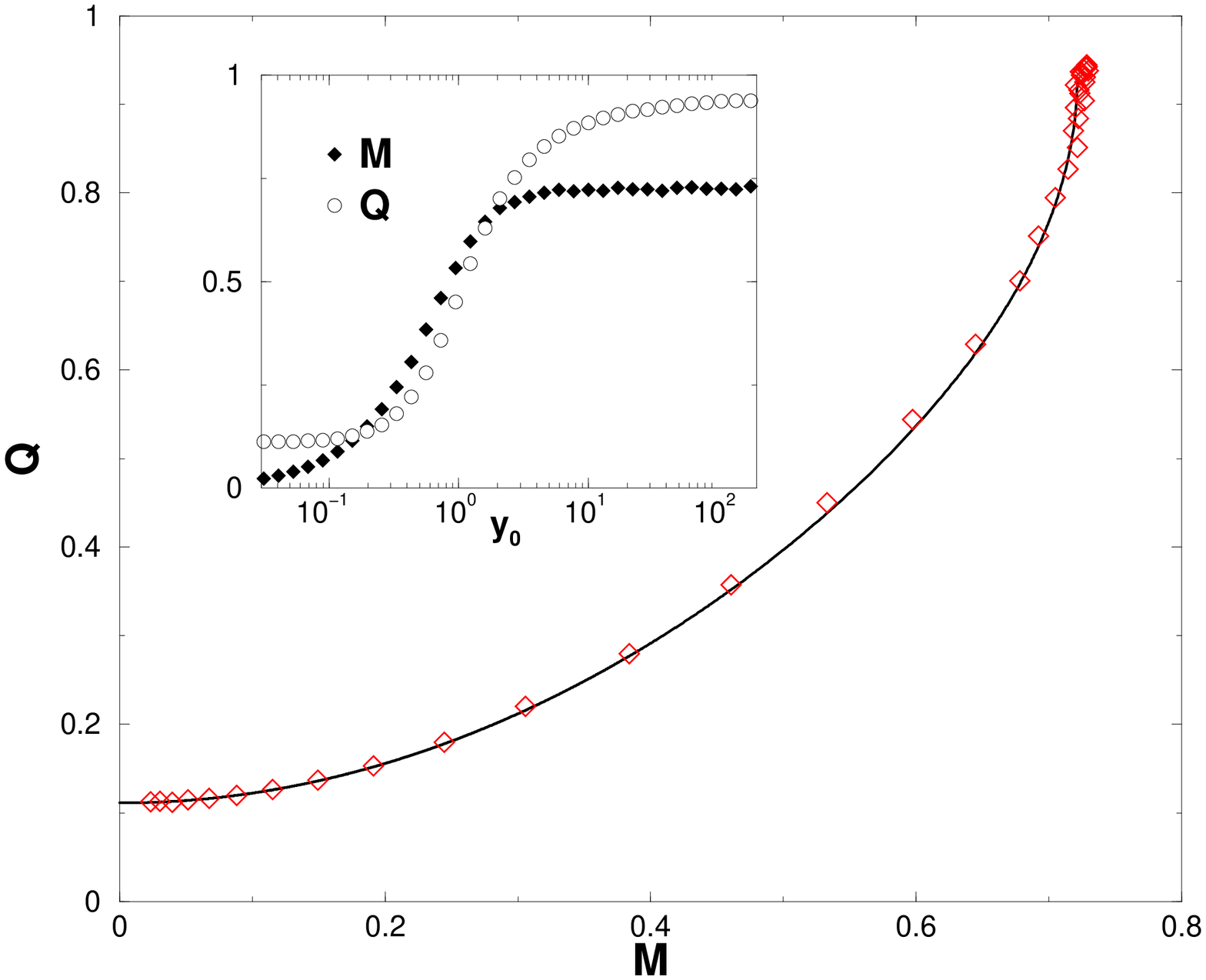,width=8cm,angle=270}}
\caption{Relation between $Q$ and $M$, for $\alpha=0.1$, derived from
analytic calculation (full line) and from numerical simulations of the
MG with different initial conditions $y_0$ ($\diamond$, $P=32,~N=320$,
$\Gamma=0.1$). The inset shows the dependence of $Q$ and $M$ from the
initial condition $y_0$.}
\label{figQM}
\end{figure}

Which SS is reached from a particular initial condition is, of course,
a quite complex issue which requires the integration of the dynamics.
However, the relation between $Q$ and $M$ derived analytically from
Eqs. (\ref{sad1},\ref{eqM1}) can easily be checked by numerical
simulations of the MG. Figure \ref{figQM} shows that the self-overlap
$Q$ and the magnetization $M$ computed in numerical simulations with
initial conditions $y_i(0)=y_0$ for all $i$, perfectly match the
analytic results. The inset of this figure shows how the final
magnetization $M$ and the self-overlap $Q$ in the SS depend on the
asymmetry $y_0$ of initial conditions.

\begin{figure}
\centerline{\psfig{file=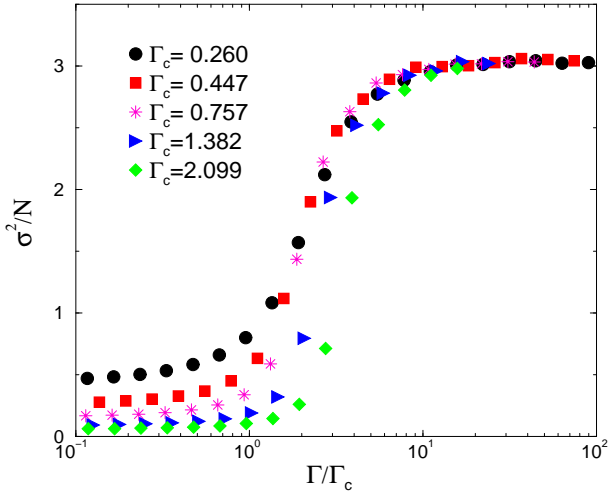,width=8cm,angle=0}}
\caption{Global efficiency $\sigma^2/N$ versus $\Gamma/\Gamma_c$ for
$\alpha=0.1<\alpha_c$, $N=160$ agents and different initial conditions
$y_0$. The value $\Gamma_c$ is computed from Eq. \req{Gammac} and is 
shown in the legend.}
\label{figs2gq}
\end{figure}

Let us finally discuss the dependence on $\Gamma$ for asymmetric
initial conditions. Eq. \req{Gammac} provides a characteristic value
of $\Gamma$ as a function of $\alpha$ and $Q$. This theoretical
prediction is tested against numerical simulations of the MG in
Fig. \ref{figs2gq}: when plotted against $\Gamma/\Gamma_c$, the curves
of $\sigma^2/N$ obtained from numerical simulations, approximately
collapse one onto the other in the large $\Gamma$ region.
Fig. \ref{figs2gq} suggests that $\Gamma_c$ in Eq. \req{Gammac}
provides a close lower bound for the onset of saturation to a constant
$\sigma^2$ for large $\Gamma$. We find it remarkable that a formula
such as Eq. \req{Gammac} which is computed in the limit $\Gamma\to 0$,
is able to predict the large $\Gamma$ behavior.

With respect to the dependence on initial conditions, we observe that
$\Gamma_c$ is an increasing function of $Q$ and hence it increases
with the asymmetry $y_0$ of initial conditions.  Hence, for a fixed
$\Gamma$, the fluctuation dependent part $\Sigma$ of $\sigma^2$
decreases with $y_0$ because it is an increasing function of
$\Gamma/\Gamma_c$. This effects adds up to the decrease of the
$\Gamma$ independent part of $\sigma^2$ discussed previously. 

Fig. \ref{figs2gq} also shows that the $\Gamma\gg\Gamma_c$ state is
independent of initial conditions $y_0$. This can naively be understood
observing that stochastic fluctuations induce fluctuations $\delta
y_i$ which increase with $\Gamma$. For $\Gamma\gg 1$ the asymmetry
$y_0$ of initial conditions is small compared to stochastic
fluctuations $\delta y_i$ and hence the system behaves as if
$y_0\approx 0$. 

\subsection{The maximally magnetized stationary state}

\begin{figure}
\centerline{\psfig{file=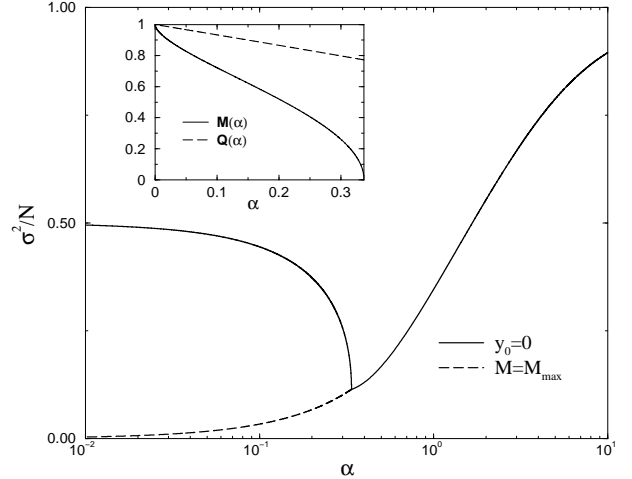,width=8cm,angle=270}}
\caption{$\sigma^2/N$ for the MG with initial conditions $y_0=0$ 
(full line) and $y_0\to\infty$ (dashed line). The inset reports 
the behavior of $M$ and $Q$ in the $y_0\to\infty$ SS. Note that 
$Q$ is linear in $\alpha$.}
\label{figs2vsaM}
\end{figure}

The maximally magnetized SS (MMSS), obtained in the limit $y_0\to\infty$,
is also the one with the largest value of 
$Q$, and hence with the smallest value of $\sigma^2=N(1-Q)/2$. 
$\sigma^2/N$ is plotted against $\alpha$ both for symmetric
$y_0=0$ initial conditions and for maximally asymmetric ones 
$y_0\to\infty$ in fig. \ref{figs2vsaM}. The inset shows the behavior 
of $Q$ and $M$ in the MMSS.

Remarkably we find that $\sigma^2/N$ vanishes linearly with $\alpha$
in the MMSS\footnote{This results was also found analytically in \cite{London}}. This means that, at fixed $P$, as $N$ increases the
fluctuation $\sigma^2$ remains constant. This contrast with what
happens in the $y_0=0$ state, for $\Gamma\ll\Gamma_c$, where
$\sigma^2$ increases linearly with $N$, and with the case $\Gamma\gg
\Gamma_c$ where $\sigma^2\propto N^2$ \cite{Savit,MC}. Note also that
the lowest curve of Fig. \ref{figs2vsaM} also gives an upper bound to
the $\sigma^2$ of Nash equilibria (see Refs. \cite{CMZ,MCZ,Andemar}).

The MMSS is also the most stable state against crowd effects: if we
put $Q(\alpha,y_0=\infty)\cong 1-c\alpha$, as appropriate for the MMSS
we find that $\Gamma_c\sim 1/\alpha$ diverges with $\alpha$.

\section{Conclusions}

We have clarified the correct derivation of continuous time dynamics
for the MG. This on the one hand reconciles the two current approaches
\cite{CMZ,Oxf2}. On the other it leads to a complete understanding of
the collective behavior of the MG. We confirm that stationary states
are characterized by the minimum of a Hamiltonian which measures the
predictability of the game. For $\alpha>\alpha_c$ we find a complete
analytic solution, whereas for $\alpha<\alpha_c$ the statistical
mechanics approach of Ref. \cite{CMZ} is valid for $\Gamma\to 0$. It
is in principle possible to introduce the new elements discussed here
in the approach of Ref. \cite{CMZ} and to derive a full analytic
solution. We have indeed derived the first term of the series
expansion for $\Gamma\ll 1$, which agrees perfectly with numerical
data. The extension of the approach of Ref. \cite{CMZ} involves 
lengthy calculations and it shall be pursued elsewhere.

Finally we note that the results derived in this paper generalize to
more complex models. It is worth to remark that the solution to the FP
equation is no more factorizable, in general, when agents account for
their market impact as in Refs. \cite{CMZ,MC,MCZ}.
Hence, as long as there are unfrozen agents, we expect that the
stationary state depends on $\Gamma$. However, when the agents take fully into
account their market impact, all of them
are frozen and the conclusion that agents converge to Nash equilibria 
remains valid.

We acknowledge constructive discussions with A.C.C. Coolen,
A. Engel, J.P. Garrahan, J.A.F. Heimel and D. Sherrington.

\appendix

\section{Non-linear minority games}

Take a generic dynamics
\[
U_{s,i}(t+1)=U_{s,i}(t)-a_{s,i}^{\mu(t)}g\left[A(t)\right]
\]
with $g(x)$ some function. When we carry out the limit to continuous
time we find a deterministic term which is proportional to
$-\ovl{a_{s,i}\avg{g(A)}}$. The stationary state conditions then read
\[
v_i=-\ovl{a_{s,i}\avg{g(A)}},~~~\hbox{if $f_{s,i}>0$}
\]
and 
\[
v_i>-\ovl{a_{s,i}\avg{g(A)}},~~~\hbox{if $f_{s,i}=0$}.
\]

For any fixed $\mu$, $A(t)$ is well
approximated by a Gaussian variable with mean 
\[
\avg{A|\mu}=\sum_{i,s}f_{s,i}a_{s,i}^\mu
\] 
and variance $D=\sigma^2-H$. Here we neglect dependences on
$\mu$. Also we treat $D$ as a parameter and neglect its dependence 
on the stationary state probabilities $f_{s,i}$.  Hence
we can write
\[
\Avg{g(A)|\mu}=\int_{-\infty}^\infty\frac{dx}{\sqrt{2\pi}}e^{-x^2/2}
g\left(\avg{A|\mu}+\sqrt{D}x\right)
\]
The stationary state conditions of the dynamics above can again be
written as a minimization problem of the functional

\[
H_g=\frac{1}{P}\sum_{\mu=1}^P\int_{-\infty}^\infty\frac{dx}{\sqrt{2\pi}}e^{-x^2/2}
G\left(\sum_{i,s}f_{s,i}a_{s,i}^\mu+\sqrt{D}x\right)
\]
with
\[
g(x)=\frac{dG(x)}{dx}
\]
and $D=\sigma^2-H$ which must be determined self-consistently.

Indeed taking the derivative of $H_g$ w.r.t. $f_{s,i}$ and imposing
the constraint $f_{s,i}\ge 0$ and normalization, we arrive at exactly
the same equations which describe the stationary state of the
process. 

The Hamiltonian for the original MG is derived setting $g(x)={\rm
sign}\,x$, which leads to

\[
H_{\rm sign}=\frac{1}{P}\sum_{\mu=1}^P\left[\frac{1}{\sqrt{\pi}}
e^{-\avg{A|\mu}^2/D}+\frac{\avg{A|\mu}}{\sqrt{D}}
{\rm erf}\left(\frac{\avg{A|\mu}}{\sqrt{D}}\right)\right].
\]

The analysis of stochastic fluctuations can be extended to non-linear
cases in a straightforward manner. Again the key point is that the
dynamics is constrained to the linear space spanned by the vectors
$\ket{a^\mu}$. For $\alpha>\alpha_c$ we have no dependence on initial
conditions. However it is not easy to show in general that the
distribution of scores factorizes across agents. This means that there
may be a contribution of fluctuations to $\sigma^2$ -- i.e. $\Sigma>0$
-- so we cannot rule out a dependence of $\sigma^2$ on
$\Gamma$. Numerical simulations for $g(x)=\sign x$ show that such a
dependence, if it exists, is very weak. Anyway even though $\sigma^2$
only depend on $f_{s,i}$, the minimization problem depends on
$D=\sigma^2-H$ which must then be determined self-consistently.

For $\alpha<\alpha_c$ the dependence on initial conditions induces a
correlation of scores across agents. As a result $\sigma^2$ depends on
$\Gamma$ just as in the linear case discussed above.

\section{Small $\Gamma$ expansion}

For $\Gamma\ll 1$ it is appropriate to consider $\beta\gg 1$ and to 
take
\[
y_i={\rm arc}\tanh m_i+\frac{z_i}{\sqrt{\beta}}
\]
so that $\beta[\log\cosh y_i-m_iy_i]\simeq 
\frac{1}{2}(1-m_i^2)z_i^2+O(\beta^{-1/2})$. Hence we have to sample 
a distribution
\[
P\{z_i\}\propto e^{-\frac{1}{2}\sum_i(1-m_i^2)z_i^2}
\]
where $z_i$ has the form
\[
z_i=\sum_{\mu=1}^P c^\mu\xi_i^\mu.
\]
It is convenient to express everything in terms of the coefficients
$c^\mu$. Their pdf is derived from that of $z_i$ and it reads:
\[
P\{c^\mu\}\propto e^{-\frac{1}{2}\sum_{\mu,\nu}c^\mu
T^{\mu,\nu}c^\nu},
~~~T^{\mu,\nu}=\sum_{i=1}^N(1-m_i^2)\xi_i^\mu\xi_i^\nu.
\]
From this we find $\avg{c^\mu c^\nu}=[T^{-1}]^{\mu,\nu}$.

Now we split the term $\Sigma(\beta)$ in two contributions,
\beas
\Sigma(\beta)&=&\ovl{\avg{\left[\sum_{i=1}^N\xi_i(\tanh
y_i-m_i)\right]^2}}+\\
&~&\sum_{i=1}^N\ovl{\xi_i^2}\left[m_i^2-(\tanh
y_i)^2\right]
\eeas
and work them out separately. For the first we use
\[
\sum_{i=1}^N\xi_i^\mu(\tanh y_i-m_i)=\frac{1}{\sqrt{\beta}}
\sum_\nu T^{\mu,\nu}c^\nu
\]
so that 
\beas
\ovl{\avg{\left[\sum_{i=1}^N\xi_i(\tanh
y_i-m_i)\right]^2}}&=&\frac{1}{\beta P}\sum_{\mu,\nu,\gamma}
T^{\mu,\nu}T^{\mu,\gamma}\avg{c^\nu c^\gamma}\\
&=&
\frac{{\rm Tr} T}{\beta P}\cong \frac{1-Q}{2\beta}N.
\eeas

Within the approximation $(1-3m_i^2)\approx (1-3Q)$ we are
able to derive a closed expression also for the second term:
\beas
\sum_{i=1}^N\ovl{\xi_i^2}\left[m_i^2\!-\!(\tanh
y_i)^2\right]&=&\frac{1}{\beta}\sum_{i=1}^N\ovl{\xi_i^2}
(1\!-\!m_i^2)(1\!-\!3 m_i^2)\avg{z_i^2}\\
&\approx & \frac{\alpha}{\beta}\frac{1-3Q}{2}N.
\eeas

Hence we find
\[
\Sigma(\beta)\cong\left[\frac{1-Q}{2}+\frac{1-3Q}{2}\alpha\right]
\frac{1}{\beta}+O(\beta^{-2}).
\]
This and Eq. \req{beta} lead to Eq. \req{s2approx}.


\begin{thebibliography}{99}
\bibitem{NotaN} Indeed
$a_{s,i}^\mu \avg{A|\mu}_\pi$ is of order $\sqrt{N}$ but its sign
fluctuates. When we average over $P\sim N$ different $\mu$, we get a
quantity of order $N^0$. We are implicitly assuming that 
$\ovl{\avg{A}}\simeq 0$ which definitely holds for large times.
\end{thebibliography}
\end{document}